\documentstyle[12pt,epsf,epsfig]{article}
\def\lpmb#1{\mbox{\boldmath $#1$}}
\newcommand{\bp}{{\bf p}}
\newcommand{\bk}{{\bf k}}

\newcommand{\btab}{\begin{tabbing}}
\newcommand{\etab}{\end{tabbing}}

\newcommand{\beqn}{\begin{equation}}
\newcommand{\eeqn}{\end{equation}}
\newcommand{\barr}[1]{\begin{array}{#1}}
\newcommand{\earr}{\end{array}}
\newcommand{\beqna}{\begin{eqnarray}}
\newcommand{\eeqna}{\end{eqnarray}}
\newcommand{\btablec}{\begin{table} \begin{center}}
\newcommand{\etablec}{\end{center} \end{table}}

\newcommand{\gapproxeq}{\lower.7ex\hbox{$\;\stackrel{\textstyle>}
{\sim}\;$}}
\newcommand{\plabel}[1]{\label{#1}}
\newcommand{\pbibitem}[1]{\bibitem{#1}}
\def\question#1{}
\input epsf

\begin{document}
\title{
\begin{flushright} 
\small{hep-ph/0304231} \\ 
\small{LA-UR-03-2566}  
\end{flushright} 
\vspace{0.6cm}  
\Large\bf Using $\pi_2(1670)\rightarrow 
b_1(1235)\:\pi$ to constrain hadronic models}
\vskip 0.2 in
\author{Philip R. Page\thanks{\small \em E-mail: prp@lanl.gov} \\
{\small \em Theoretical Division, MS B283, Los Alamos
National Laboratory,}\\ 
{\small \em  Los Alamos, NM 87545, U.S.A.} \\
Simon Capstick\thanks{\small \em E-mail: capstick@csit.fsu.edu} \\
{\small \em Department of Physics, Florida State University,
Tallahassee, FL 32306-4350, U.S.A.}}
\date{}
\maketitle
\begin{abstract}{
We show that current analyses of experimental data indicate that the
strong decay mode $\pi_2\rightarrow b_1\pi$ is anomalously small.
Non-relativistic quark models with spin-1 quark pair
creation, such as $^3P_0$, $^3S_1$ and $^3D_1$
models, as well as instanton and lowest order one-boson (in this case
$\pi$) emission models, can accommodate the analyses of experimental
data, because of a quark-spin selection rule.  Models and effects that
violate this selection rule, such as higher order one-boson emission
models, as well as mixing with other Fock states, 
may be constrained by the small $\pi_2\rightarrow b_1\pi$
decay. This can provide a viability check on newly proposed decay
mechanisms. We show that for mesons made up of a heavy quark and
anti-quark, the 
selection rule is exact to all orders of Quantum Chromodynamics
(QCD) perturbation theory.}
\end{abstract}
\bigskip

PACS number(s): 13.25.Jx \hspace{.2cm}14.40.Cs
\hspace{.2cm} 12.39.Ki \hspace{.2cm} 12.38.Bx


\section{Analyses of experimental data on $\pi_2(1670)\rightarrow b_1(1235)\:\pi$}

Recently, the VES Collaboration published for the first time an upper
bound of $0.0019$ on the branching fraction for Br$[\pi_2\rightarrow
b_1\pi]$, at the $97.7$\% confidence level. This branching fraction is
measured in $37$ GeV $\pi^-$ collisions on a nucleus, in the reaction
$\pi^- A \rightarrow \omega\pi^-\pi^0 A^\ast$~\cite{amelin}.  This
small branching fraction is consistent with a preliminary analysis
performed by the E852 Collaboration~\cite{popov} of data on the
reaction $\pi^- p \rightarrow \omega\pi^-\pi^0 p$, in collisions of an
$18$ GeV $\pi^-$ beam with a proton target.

\begin{table}[t]
\begin{center}
\caption{ Branching fractions, and ratios $R(X) = |{\cal M}(X)|^2\; /
\; |{\cal M}(f_2\pi)|^2$ and $\tilde{R}(X) = |\tilde{\cal M}(X)|^2\; /
\; |\tilde{\cal M}(f_2\pi)|^2$ of partial widths with phase space and flavor
factors removed to those of the dominant decay mode. 
${\cal M}$ and $\tilde{\cal M}$ are defined in the text. The decay is 
assumed to proceed via the bold-faced $L$ wave, since in all modes
[except for $f_2\pi$, where~\protect\cite{pdg00} the $D$ wave is
$(0.18\pm 0.06)^2=(3.2\pm 2.2)\%$ of the $S$ wave] the contributions
from the different partial waves are not known. Although the branching
fractions do not add to unity, since Ref.~\protect\cite{pdg00}
constrained a subset of these modes by unitarity, those outside of
this subset were defined relative to the dominant $f_2\pi$ mode, and
so this does not affect the ratios $R(X)$ and $\tilde{R}(X)$.
The constraint for the $\rho(1450)\pi$ mode is incorrectly 
quoted~\protect\cite{dorofeev} in 
Refs.~\protect\cite{amelin,pdg00} and should read $Br[\pi_2(1670)\rightarrow
 \rho(1450)\pi]\; Br[\rho(1450)\rightarrow \omega\pi] < 0.36 \%$.
Since $Br[\rho(1450)\rightarrow \omega\pi]$ is poorly known, 
estimates for a branching ratio of a third are provided.
\plabel{wid}}
\begin{tabular}{|c|l|l|c|c||r|r|}
\hline 
Mode $X$ & $p$ (GeV) & $L$ & $f^2$ & Br$(\pi_2\rightarrow X)$ 
(\%)~\protect\cite{pdg00} &
$R(X)$ & $\tilde{R}(X)$\\
\hline \hline 
$f_2\pi$       & 0.326 & {\bf S}, D, G & 2 &$56.2\pm 3.2$&  1.00  &  1.00 \\
$\sigma\pi$    & 0.634 & {\bf D}       & 2 &$13\pm 6$    &  0.73  &  1.00 \\
$\omega\rho$   & 0.308 & {\bf P}, F    & 2 &$ 2.7\pm 1.1$&  0.53  &  0.53 \\
$\rho(1450)\pi$& 0.143 & {\bf P}, F    & 4 &$<0.36\times 3$ &$<0.36\times 3$&$<0.33\times 3$\\
$\rho\pi$      & 0.649 & {\bf P}, F    & 4 &$31\pm 4$     &  0.33 &  0.46 \\
$K\bar{K}^\ast$& 0.450 & {\bf P}, F    & 2 &$ 4.2\pm 1.4$ &  0.27 &  0.30 \\
$b_1\pi$       & 0.363 & {\bf D}       & 4 &$<0.19$       &$<0.09$&$<0.09$\\
\hline 
\end{tabular}
\end{center}
\end{table}

The decay $\pi_2\rightarrow b_1\pi$ is allowed by conservation of
parity, angular momentum, isospin and G-parity, and so its strength
should be comparable with that of other decays which are allowed by
the same quantum numbers, which are conserved to an extraordinary
degree by the strong interactions. In order to show that the branching
ratio is small for dynamical reasons, independent of any model,
factors due to phase space and flavor should be removed. The standard
expression for the partial width is~\cite{pdg00}
\beqn
\Gamma = \frac{p}{8\pi \; (2 J_{\pi_2} +1) \; 
m_{\pi_2}^2}\; |p^L f {\cal M}|^2
\eeqn
where $m_{\pi_2}$ and $J_{\pi_2}$ are the mass and total angular momentum of
the decaying $\pi_2$, the decay momentum $p$ is measured in the rest
frame of the $\pi_2$, the relative orbital angular momentum of the decay
products is $L$, and $p^L f {\cal M}$ is the decay amplitude. The
amplitude with the phase space ($p^L$) and flavor ($f$) factors
removed is ${\cal M}$. In Table~\ref{wid} we show the ratios of
$|{\cal M}|^2$ for the observed decay modes of the $\pi_2$ to that of
the dominant decay mode ($f_2\pi$).  A further refinement is to remove
the dependence on the kinematics of the decays from the form
factors of the initial and final mesons. With universal Gaussian wave functions for the mesons, this can be accomplished by defining ${\cal M} = \exp
(-p^2/[12\beta^2])\:\tilde{\cal M}$, where $\beta =
0.4$ GeV~\cite{d2637}. 

The ratios of the squares of these amplitudes with the flavor, phase
space, and kinematic factors removed is also shown in
Table~\ref{wid}.  It is evident that the $b_1\pi$ decay is a factor of
between 3 and 11 weaker than the other decay modes for dynamical
reasons, making it anomalously small. This is emphasized by
Fig.~\ref{merc}, which shows the $|{\cal M}|^2$ ratios plotted
logarithmically. Since there is only an experimental upper bound on
the $b_1\pi$ mode, this suppression factor could be even larger. There
is also evidence from recent analyses of E852 data~\cite{KuhnEugenio}
of a $\pi_2(1670)$ signal in the $f_1\pi$ and $a_2\eta$ final
states. The discovery of additional final states will have the effect
of further reducing the $b_1\pi$ branching fraction. We urge future
experiments to put more restrictive bounds on the $\pi_2\to b_1\pi$
decay mode.

\begin{figure}[t]
\begin{center}
\epsfig{file=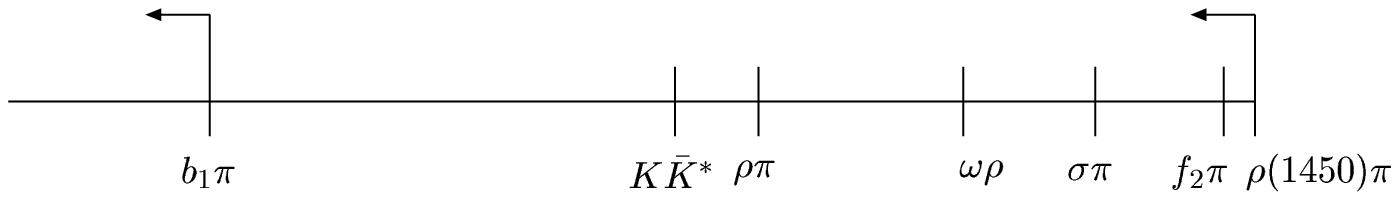,width=1.0\linewidth,clip=}
\end{center}
\vspace{-1.0cm}\caption{\plabel{merc} Ratios
($|{\cal M}(X)|^2\; / \; |{\cal M}(f_2\pi)|^2$ ) plotted logarithmically.}
\end{figure}

\section{Models that can accommodate $\pi_2(1670)\rightarrow b_1(1235)\:\pi$}

The decay $\pi_2\rightarrow b_1\pi$ is particularly clean in the sense
that it is only sensitive to OZI allowed decays.  This is because
OZI-forbidden decay processes, which allow the creation of either the
isovector $\pi_2$, $b_1$ or $\pi$ out of isoscalar gluons, are 
forbidden by isospin symmetry (see Fig. \ref{OZI}). 
The suppression of isospin symmetry breaking amplitudes is
much greater than that of OZI forbidden amplitudes, the latter being
about a factor of 10.
\begin{figure}[t]
\begin{center}
\epsfig{file=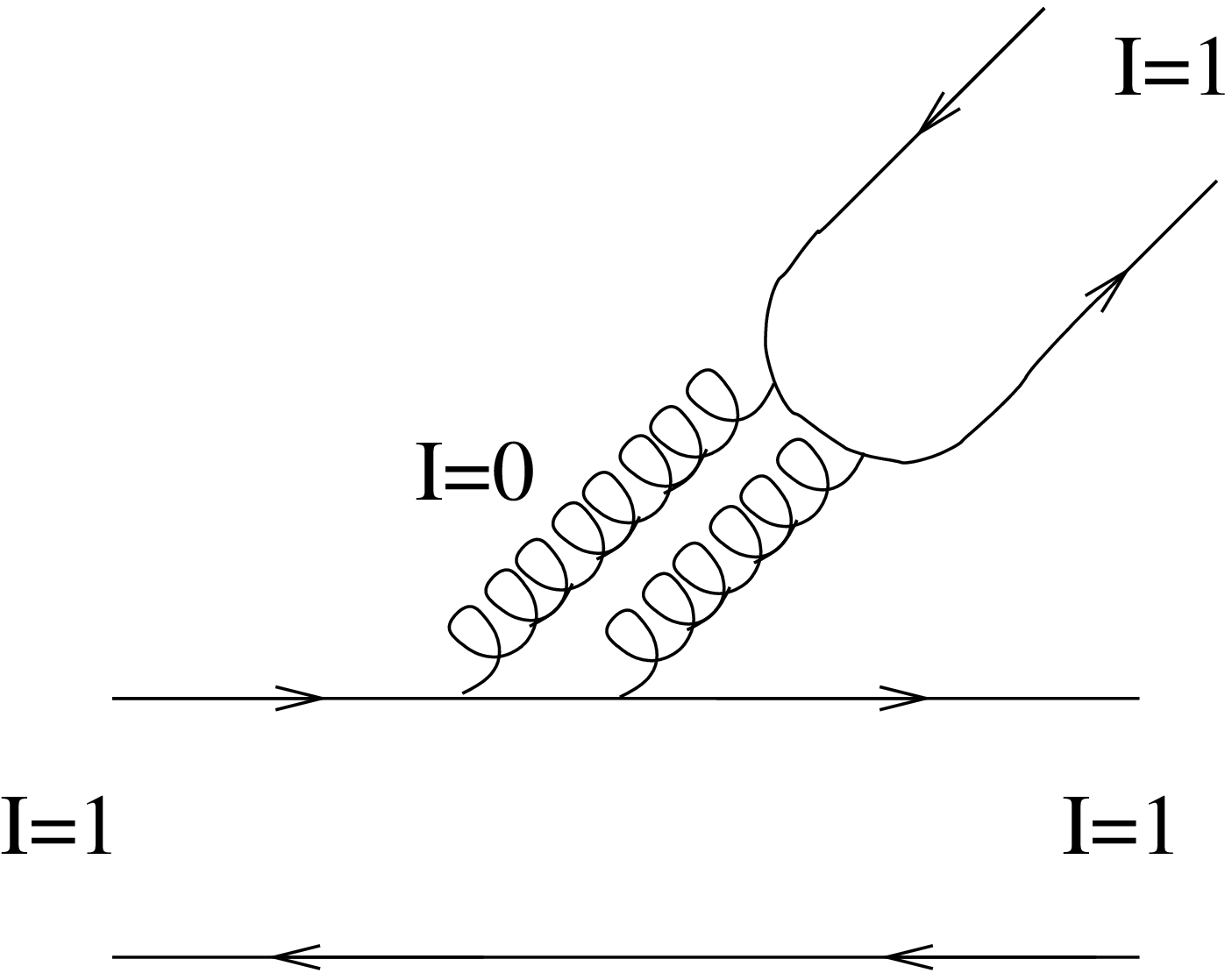,width=0.3\linewidth,clip=}
\end{center}
\vspace{-0.5cm}\caption{\plabel{OZI} OZI forbidden decays of an isovector meson to a
pair of isovector mesons.}
\end{figure}

In non-relativistic quark-pair-creation models, where OZI-allowed
meson decay processes are modeled by an initial $q\bar{q}^\prime$ pair
decaying to the two pairs $q\bar{q}^{\prime\prime}$ and
$q^{\prime\prime}\bar{q}^\prime$ (see Fig.~\ref{diag}), 
a simple selection rule arises when
all the mesons have quark-spin $S=0$.  If the
$q^{\prime\prime}\bar{q}^{\prime\prime}$ pair is created with
quark-spin $S_{\mbox{\small pair}}=1$, then conservation of quark-spin implies that the
amplitude is zero~\cite{close,barnes}. In the quark model, conventional mesons with
$S=0$ have $J^{PC} = 0^{-+},\; 1^{+-},\; 2^{-+},\;
3^{+-},\; 4^{-+},\; 5^{+-}, \ldots$, of which only states
corresponding to the first three $J^{PC}$ have been established
experimentally~\cite{pdg00}. The isovector resonances with these three
$J^{PC}$ and in their radial ground states are $\pi, b_1$ and $\pi_2$,
respectively. The only kinematically allowed decay involving these
three $S=0$ resonances is $\pi_2\rightarrow b_1\pi$. Moreover, all other
kinematically allowed decays involving $\pi, b_1, \pi_2$, and their
isoscalar partners, are forbidden by the quantum numbers conserved by
the strong interaction. The first explicit mention of the quark-spin
selection rule or its application to $\pi_2\rightarrow b_1\pi$
was in Refs.~\cite{close}, although it is implicit in Ref.~\cite{pene}.

No other strong decay involving conventional mesons composed of quarks
other than $u,d$ quarks currently appears to be able to test the
selection rule. Decays $q\bar{q}\rightarrow q\bar{q}+q\bar{q}$ with
$q \in \{ {s,c,b} \}$, where each meson is in its radial ground state with
the $S=0$ quantum numbers $J^{PC} = 0^{-+},\; 1^{+-}$ or $2^{-+}$, are
forbidden for the same reasons as decays between the isoscalar
resonances above. With the exception of the pseudoscalars, quark-model
mesons with
the open flavor structure $K,\; D,\; D_s,\; B,\; B_s$ or $B_c$, and
lying on the un-natural parity sequence $J^P=0^-,\; 1^+,\; 2^-,\; 3^+,
\; 4^-,\; 5^+,\ldots$ are mixtures of $S=0$ and $S=1$ states, since
$S=1$ components are no longer excluded by charge conjugation symmetry.  
In QCD, if one of the initial or final mesons in the decay has this
open flavor structure, a second meson must also. This implies that the
selection rule can only be tested in decays involving open flavor
mesons if there are two open flavor pseudoscalar mesons
involved. Since two pseudoscalars with an arbitrary relative angular
momentum couple to the natural-parity sequence $J^P=0^+,\; 1^-,\;
2^+,\; 3^-, \; 4^+,\; 5^-,\ldots$, the $S=0$ selection rule cannot be
tested with decays involving open flavor mesons. It is, therefore,
evident how central and unique the decay mode $\pi_2\rightarrow
b_1\pi$ is for testing this selection rule.
\begin{figure}[t]
\begin{center}
\vspace{-1cm}\epsfig{file=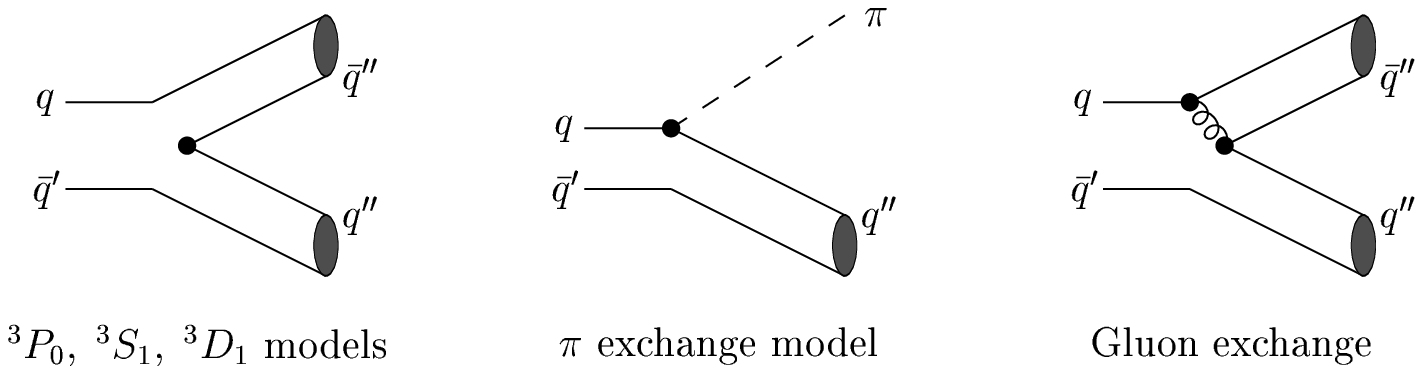,width=1.0\linewidth,clip=}
\end{center}
\vspace{-1.0cm}\caption{\plabel{diag} The OZI allowed decay of an initial meson to two
final mesons in various models.}
\end{figure}

The selection rule obtains when the $\pi_2,\; b_1$ and $\pi$ are
treated non-relativistically as $S=0$ mesons in the quark model. Remarkably,
relativistic interactions cannot introduce $S=1$ components
in the $\pi_2$, $b_1$ and $\pi$ wave functions, so that the
selection rule remains valid after relativistic interaction corrections
to the quark model.
This is because the $q\bar{q}'$ Fock state
wave function of the $\pi_2$ can only have $^1D_2$
quantum numbers before relativistic interactions, and the interactions
cannot change that. The analogous argument holds for $b_1$ and $\pi$.
Even in the fully relativistic equal-time Bethe-Salpeter equation
the selection rule is exact~\cite{ricken}.
It remains an open question whether a selection rule would be found
in field theoretic calculations of
$\pi_2\rightarrow b_1\pi$, e.g. in the lattice QCD,
QCD sum rule, and Dyson-Schwinger equation approaches.

It has been pointed out that a success of the non-relativistic $^3P_0$
pair-creation model (Fig.~\ref{diag}), where 
$S_{\mbox{\small pair}}=1$, is the fact
that the decay $\pi_2\rightarrow b_1\pi$ is predicted to
vanish~\cite{barnes}.  Other decay models where 
$S_{\mbox{\small pair}}=1$, such as the non-relativistic
chromo-electric string-breaking model where the pair has $^3S_1$ or
$^3D_1$ quantum numbers~\cite{alcock} (Fig.~\ref{diag}), will also have this
suppression. Both the $^3P_0$ and $^3S_1$ models involve a decay
operator proportional to $\lpmb{\sigma} \cdot \bp$, where the
$\lpmb{\sigma}$ is the spin of the created quark anti-quark pair, and
$\bp$ is a momentum operator. It is not surprising that the $^3P_0$,
$^3S_1$ and $^3D_1$ models obey the selection rule, since these all
treat the quarks non-relativistically, as though they are heavy. This
is a special case of a result that is shown in Appendix~\ref{app}:
when each of the mesons participating in the decay is composed of a
very heavy quark and anti-quark, the selection rule is exact to all
orders of QCD perturbation theory.

Since 't Hooft's instanton-induced six-quark vertices only affects strong 
decays where all participating mesons have J=0, and their singlet flavor 
structure requires the presence of a strange quark (and anti-quark), decay 
models based on these vertices also predict vanishing $\pi_2\to b_1\pi$ 
decay~\cite{munz}.


%
\section{One-boson emission models}

The one-boson exchange (OBE) model describes the coarse features of the
baryon spectrum as being due to confinement and the exchange of
pseudoscalar~\cite{GR} and scalar and vector~\cite{Graz} bosons
between the quarks. For light-quark baryons an important pseudoscalar
exchange potential comes from pion exchange. This model is not applied
to meson spectroscopy. Two reasons are often given for this. The first
is that if the light pseudoscalar bosons are the
pseudo-Goldstone bosons of spontaneously-broken chiral symmetry, then
it is inconsistent to also treat them as quark-anti-quark bound states and
allow OBE to act between the quark and anti-quark. This argument would
not appear to be applicable to heavier quark-anti-quark bound states
such as the $\pi_2(1670)$ and $b_1(1235)$.

A second reason for not applying this model to the meson spectrum is
that if one-boson-exchange in baryons is viewed microscopically, with
the pion treated as a $q\bar{q}$ pair, an exchange of quarks in the
process $q q^\prime \rightarrow q^\prime q$ can be viewed in one time
ordering as an exchange of $q\bar{q}^\prime$, which can be identified
with a meson.  In a meson the exchange of quarks occurs in the process
$q \bar{q}^\prime \rightarrow \bar{q}^\prime q$, which in one time
ordering is the exchange of a di-quark $q q^\prime$ and not a meson.
Exchange of mesons like the pion between quarks is, therefore, not
expected to be important to the structure of mesons, even if it is
important for baryons.

Once one admits a quark-pseudoscalar meson vertex as employed in baryon
spectroscopy, this vertex naturally leads baryons to decay to a
baryon plus a pseudoscalar meson, and mesons to decay to a meson and a
pseudoscalar meson. For this reason the OBE model of
baryon spectroscopy {\it implies} a one-boson emission decay
model in baryons and in spatially excited mesons. This model should,
therefore, be confronted with $\pi_2\rightarrow b_1\pi$.

In the $^3P_0$, $^3S_1$ and $^3D_1$ models,
pionic decay of mesons proceeds via
$q\bar{q}^\prime$ pair decaying to the two final meson pairs
$q\bar{q}^{\prime\prime}$ and $q^{\prime\prime}\bar{q}^\prime$, one of
which is identified with the pseudoscalar boson. As shown in 
Fig.~\ref{diag}, the one-pion emission
model has either $q\rightarrow q^{\prime\prime}\pi$, or
$\bar{q}^\prime\rightarrow\bar{q}^{\prime\prime}\pi$.  The
lowest order one-pion coupling to the quark or anti-quark is given by
the Lagrangian density~\cite{Becchi,Hen,Goity}
\beqn {\cal L}_\pi = i \frac{g_A^q}{2f_{\pi}}
\bar{\psi}(x)\gamma_5\gamma_{\mu}
\partial^{\mu}\vec{\pi}(x)\cdot\vec{\tau}\psi(x) + \mbox{H.c.}  \eeqn
An expansion of this axial current gives a decay operator of the form
$\lpmb{\sigma}_q \cdot \bk$ (Eqs. 2 and 28 of Ref.~\cite{Hen}), where
$\lpmb{\sigma}_q$ is the spin of the quark emitting the pion, and
$\bk$ is the pion momentum.  This means that the operator creating the
boson is a vector operator in the space of the spin of the decaying
meson, and so cannot link an initial $S=0$ meson to a final $S=0$
meson, so the selection rule is also valid for lowest order one-boson
emission.

We conclude that the phenomenologically successful pair-creation model
for light-light mesons (the $^3P_0$ model)~\cite{barnes}, the
chromo-electric string-breaking model ($^3S_1$ or $^3D_1$ model), 
instantons~\cite{munz}, and
the lowest order one-boson emission model, which has successfully been
applied to the decay of heavy-light
mesons~\cite{Hen,Goity}, are consistent with the experimental decay width of 
$\pi_2\rightarrow b_1\pi$. 

\section{Models possibly constrained by $\pi_2(1670)\rightarrow b_1(1235)
\:\pi$}

Higher order contributions in one-boson emission models contain terms that
are not of the form $\lpmb{\sigma}_q \cdot \bp$, which violate the
selection rule. An example is interactions where {\it both} a
pseudoscalar boson is emitted, {\it and} a particle is exchanged
between the quark and anti-quark in the initial meson (Eqs.~13, 38 and
39 of Ref.~\cite{Hen}). The amplitudes corresponding to the higher
order contributions can be similar in size to those corresponding to
the lowest order contribution\footnote{See Table 4 of
Ref.~\cite{Hen}. Note that the size of the part of the higher-order
interaction that is not of the form $\lpmb{\sigma}\cdot\bp$ is not
evaluated in Ref.~\protect\cite{Hen}.}. This suggests that consistency
with the small decay branch for $\pi_2\rightarrow b_1\pi$ can
constrain models which do not obey the selection rule, such as the
higher order contributions introduced in one-pseudoscalar-boson emission
models~\cite{Hen} to cure problems with the lowest order
contribution~\cite{Hen,Goity}. It can also provide a viability check on
proposed decay mechanisms.  An example, depicted in Fig.~\ref{diag},
 is where there is a single
gluon exchanged between a quark in the decaying hadron and the vertex
at which the quark pair is created. Although this one-gluon exchange
quark pair creation decay mechanism violates the selection
rule\footnote{One-gluon exchange involves both Coulomb and transverse
interactions. The former has a simple $\lpmb{\sigma}\cdot{\bp}$ pair
creation operator, but the latter involves {\it both} spin vector pair
creation, and an additional term at the vertex where the quark or
anti-quark emits a gluon (See Eqs. B5-B7 of
Ref.~\protect\cite{ackleh}).  This additional term includes a
$\lpmb{\sigma}\cdot \bp/m$ contribution~\protect\cite{ackleh}, so that
the overall transverse gluon interaction has spin vector operators at
{\it both} interaction vertices of the gluon, giving rise to a
violation of the selection rule.}, it is found to be sub-dominant
relative to the $^3P_0$ model~\cite{ackleh}, so that it is not
expected to be constrained by $\pi_2\rightarrow b_1\pi$.
If appreciable strength for
$\pi_2\rightarrow b_1\pi$, inconsistent with experiment,
is predicted by either higher order terms present in the one-boson
emission decay mechanism, or by the one-gluon exchange pair creation
decay mechanism, one of these decay models could be ruled out. This
could distinguish between the OBE and one-gluon
exchange models of the coarse features of the light baryon spectrum.

Even though the main models commonly applied to strong decays have been
discussed, a comprehensive discussion of all proposed decay mechanisms
has not been given. Such mechanisms should be confronted with
the experimental data on $\pi_2\rightarrow b_1\pi$.

\section{Further constraints due to $\pi_2(1670)\rightarrow b_1(1235)
\:\pi$}

In addition to
aspects of the decay models discussed in the previous section, further
breaking of the selection rule can arise from
mixing with other Fock states. The mixing of mesons
participating in the decay with non-$q\bar{q}'$ Fock states is constrained
by the experimentally measured $\pi_2\rightarrow b_1\pi$ width.
Examples of such mixing are mixing between the $S=0$ meson $\pi_2$ 
and the $S=1$ hybrid $\pi_2$ meson expected nearby in mass,
and non-mesonic Fock states in the pseudo-Goldstone boson $\pi$.

\section*{Acknowledgments}

Helpful discussions with P.~Eugenio and V.A.~Dorofeev are gratefully
acknowledged. This research is supported by the U.S. Department of
Energy under contracts DE-FG02-86ER40273 (SC), and W-7405-ENG-36
(PRP).

\appendix

\section{Appendix: The quark-spin selection rule is exact for heavy 
quarks\plabel{app}}

The quark-gluon interaction in the QCD Lagrangian density 
(suppressing flavor and color) is
\beqn\plabel{lag}
{\cal L} = g \bar{\psi}(x) \gamma_{\mu} {A}^{\mu} (x) \psi(x) + \mbox{H.c.}
\eeqn
%
Second quantize the free quark fields in the usual way,
\beqn\plabel{qua}
\psi(x) = \int \frac{d^3 p}{\sqrt{(2\pi)^32E(p)}} \sum_\nu \;
\left[\: a^\nu(p) u^\nu(p)e^{ip\cdot x} 
+ b^{\nu\dagger}(p) v^\nu(p)e^{-ip\cdot x}\: \right],
\eeqn
where $a^\nu(p)$ and $b^{\nu}(p)$ are the quark and anti-quark
annihilation operators. Substituting Eq.~\ref{qua} into Eq.~\ref{lag}
yields
\beqna\plabel{big}
{\cal L} &=& g \int \frac{d^3p d^3p'}{(2\pi)^3\sqrt{(2E(p))(2E(p'))}} \;
\sum_{\nu\nu'} \;
\left[\: u^{\nu\dagger}(p) \gamma_0\gamma_\mu u^{\nu'}(p') \: {A}^{\mu} (x) 
e^{i(p'-p)\cdot x} \: a^{\nu\dagger}(p)a^{\nu'}(p') \right.  \nonumber \\ 
&+& v^{\nu\dagger}(p) \gamma_0\gamma_\mu v^{\nu'}(p') \: {A}^{\mu} (x) 
e^{i(p-p')\cdot x} \: b^{\nu}(p)b^{\nu'\dagger}(p') \nonumber \\
&+& u^{\nu\dagger}(p) \gamma_0\gamma_\mu v^{\nu'}(p') \: {A}^{\mu} (x) 
e^{-i(p+p')\cdot x} \: a^{\nu\dagger}(p)b^{\nu'\dagger}(p') \nonumber   \\ 
&+& \left. v^{\nu\dagger}(p) \gamma_0\gamma_\mu u^{\nu'}(p') \: {A}^{\mu} (x) 
e^{i(p+p')\cdot x} \: b^{\nu}(p)a^{\nu'}(p')
\:\right] + \mbox{H.c.}
\eeqna
The first and second terms describe the quark and anti-quark
interactions with the gluon field, respectively, the third term
describes creation of a quark-anti-quark pair, and the fourth term
annihilation of a quark-anti-quark pair.

In the limit of very heavy quarks
\beqn
u^\nu(p) = \sqrt{2m_Q} \left( \barr{c} \chi^\nu \\ 0 \earr \right)
\hspace{0.5cm}
v^\nu(p) = \sqrt{2m_Q} \left( \barr{c} 0 \\ \chi^\nu \earr \right),
\hspace{0.5cm} 
\eeqn
where the $\chi^\nu$ are the usual Pauli spinors. Then the first
and second terms in Eq.~\ref{big} contain
\beqn\label{forward}
u^{\nu\dagger}(p) \gamma_0\gamma_\mu u^{\nu'}(p') = 
 v^{\nu\dagger}(p) \gamma_0\gamma_\mu v^{\nu'}(p')=
2m_Q\: \chi^{\nu\dagger} \chi^{\nu'} \delta_{\mu 0} = 
  2m_Q \delta_{\nu\nu'}\delta_{\mu 0},
\eeqn
so quark-gluon and anti-quark-gluon interactions do not change the
spin of heavy quarks or anti-quarks.  The third and fourth terms in
Eq.~\ref{big} contain
\beqn\label{pair}
u^{\nu\dagger}(p) \gamma_0\gamma_\mu v^{\nu'}(p') =
v^{\nu\dagger}(p) \gamma_0\gamma_\mu u^{\nu'}(p') =
2m_Q \: \chi^{\nu\dagger}\sigma_i\chi^{\nu'}\: \delta_{\mu i}
\eeqn
where $i\in \{1,2,3\}$. Hence quark-anti-quark pair
creation and annihilation involve a spin change
described by the Pauli matrices $\sigma_i$.

The spin of a propagating heavy quark remains unchanged by quark-gluon
interactions, according to the first and second terms of the
interaction in Eq.~\ref{big}, and Eq.~\ref{forward}.  The exception to
this is when the the quark travels in a Z-graph, which corresponds to
quark-anti-quark pair creation and then annihilation via the third and
fourth terms of the interaction in Eq.~\ref{big}. However, these
Z-graphs are suppressed by powers of $1/m_Q$, so that
for very heavy quarks they do not contribute. The spin of a
propagating heavy quark remains unchanged to all orders in QCD
perturbation theory.

This implies that the spin of a quark or anti-quark is changed only
when a quark-anti-quark pair is created or annihilated, through an
operator of the form $\lpmb{\sigma}\cdot{\bf A}$ (Eqs.~\ref{big}
and~\ref{pair}).  When an initial heavy-quark meson $Q\bar{Q}^\prime$
pair undergoes an OZI allowed decay to the two final heavy-quark meson
pairs $Q\bar{Q}^{\prime\prime}$ and $Q^{\prime\prime}\bar{Q}^\prime$,
the spin is only changed when the
$Q^{\prime\prime}\bar{Q}^{\prime\prime}$ pair is created\footnote{
According to Eqs.~\protect\ref{big},~\protect\ref{forward}
and~\protect\ref{pair}, only the time-like component of the gluon field
couples to the propagating quark or anti-quark, while only the
space-like component couples in the case of quark-anti-quark pair
creation or annihilation. The time-like component can change into the
space-like component via gluon-loop diagrams in both covariant and
Coulomb gauge, and also via ghost-loop diagrams in covariant gauge.}.
Also, since the individual mesons are composed of very heavy quarks,
moving non-relativistically, they have a specific quark-spin (assuming
no accidental mixing with states nearby in mass).  It follows that the
spin selection rule is exact to all orders in QCD perturbation theory
when the mesons participating in the decay are built from very
heavy quarks and anti-quarks. Light quark loops do not change these
conclusions.  For very heavy quarks, $1/m_Q$ corrections are negligible
compared to higher order corrections in $\alpha_s$,
because $\alpha_s(m_Q)$ depends only logarithmically on $m_Q$.
\question{what does NRQCD say?}

\end{document}